\documentclass[12pt]{article}
\usepackage{eqsection,subeqnarray,indent,amsfonts,epsf}

\begin{document}

\bibliographystyle{plain}

\date{November 27, 2000 \\[2mm] Revised February 9, 2001}

\title{\vspace*{-1cm}
       Transfer Matrices and Partition-Function Zeros \\
       for Antiferromagnetic Potts Models \\[5mm]
       \large\bf II.~Extended results for square-lattice chromatic polynomial}

\author{
  \\
  {\small Jesper Lykke Jacobsen}                            \\[-0.2cm]
  {\small\it Laboratoire de Physique Th\'eorique et Mod\`eles Statistiques}
                                                            \\[-0.2cm]
  {\small\it Universit\'e Paris-Sud}                        \\[-0.2cm]
  {\small\it B\^atiment 100, F-91405 Orsay, FRANCE }        \\[-0.2cm]
  {\small\tt JACOBSEN@IPNO.IN2P3.FR}                        \\[5mm]
  {\small Jes\'us Salas}                                    \\[-0.2cm]
  {\small\it Departamento de F\'{\i}sica Te\'orica}         \\[-0.2cm]
  {\small\it Facultad de Ciencias, Universidad de Zaragoza} \\[-0.2cm]
  {\small\it Zaragoza 50009, SPAIN}                         \\[-0.2cm]
  {\small\tt JESUS@MELKWEG.UNIZAR.ES}     \\[-0.2cm]     
  {\protect\makebox[5in]{\quad}}  
  \\
}

\maketitle
\thispagestyle{empty}   

\begin{abstract}
We study the chromatic polynomials for $m\times n$ square-lattice strips, 
of width $9\leq m \leq 13$ (with periodic boundary conditions) and arbitrary 
length $n$ (with free boundary conditions). We have used a transfer matrix 
approach that allowed us also to extract the limiting curves when 
$n\rightarrow\infty$. In this limit we have also obtained the isolated limiting
points for these square-lattice strips and checked some conjectures related
to the Beraha numbers. 
\end{abstract}


\clearpage

\newcommand{\be}{\begin{equation}}
\newcommand{\ee}{\end{equation}}
\newcommand{\<}{\langle}
\renewcommand{\>}{\rangle}
\newcommand{\widebar}{\overline}
\def\reff#1{(\protect\ref{#1})}
\def\spose#1{\hbox to 0pt{#1\hss}}
\def\ltapprox{\mathrel{\spose{\lower 3pt\hbox{$\mathchar"218$}}
 \raise 2.0pt\hbox{$\mathchar"13C$}}}
\def\gtapprox{\mathrel{\spose{\lower 3pt\hbox{$\mathchar"218$}}
 \raise 2.0pt\hbox{$\mathchar"13E$}}}
\def\textprime{${}^\prime$}
\def\proof{\par\medskip\noindent{\sc Proof.\ }}
\def\qed{\hbox{\hskip 6pt\vrule width6pt height7pt depth1pt \hskip1pt}\bigskip}
\def\proofof#1{\bigskip\noindent{\sc Proof of #1.\ }}
\def\half{ {1 \over 2} }
\def\third{ {1 \over 3} }
\def\twothird{ {2 \over 3} }
\def\smfrac#1#2{\textstyle{#1\over #2}}
\def\smhalf{ \smfrac{1}{2} }
\newcommand{\real}{\mathop{\rm Re}\nolimits}
\renewcommand{\Re}{\mathop{\rm Re}\nolimits}
\newcommand{\imag}{\mathop{\rm Im}\nolimits}
\renewcommand{\Im}{\mathop{\rm Im}\nolimits}
\newcommand{\sgn}{\mathop{\rm sgn}\nolimits}
\newcommand{\tr}{\mathop{\rm tr}\nolimits}
\newcommand{\diag}{\mathop{\rm diag}\nolimits}
\newcommand{\Gal}{\mathop{\rm Gal}\nolimits}
\newcommand{\mycup}{\mathop{\cup}}
\def\hboxscript#1{ {\hbox{\scriptsize\em #1}} }
\def\zhat{ {\widehat{Z}} }
\def\phat{ {\widehat{P}} }
\def\qtilde{ {\widetilde{q}} }

\def\scra{\mathcal{A}}
\def\scrb{\mathcal{B}}
\def\scrc{\mathcal{C}}
\def\scrd{\mathcal{D}}
\def\scrf{\mathcal{F}}
\def\scrg{\mathcal{G}}
\def\scrl{\mathcal{L}}
\def\scro{\mathcal{O}}
\def\scrp{\mathcal{P}}
\def\scrq{\mathcal{Q}}
\def\scrr{\mathcal{R}}
\def\scrs{\mathcal{S}}
\def\scrt{\mathcal{T}}
\def\scrv{\mathcal{V}}
\def\scrz{\mathcal{Z}}

\def\q{{\sf q}}

\def\Z{{\mathbb Z}}
\def\R{{\mathbb R}}
\def\C{{\mathbb C}}
\def\Q{{\mathbb Q}}

\def\T{{\mathsf T}}
\def\H{{\mathsf H}}
\def\V{{\mathsf V}}
\def\D{{\mathsf D}}
\def\J{{\mathsf J}}
\def\P{{\mathsf P}}
\def\QQ{{\mathsf Q}}
\def\RR{{\mathsf R}}

\def\bsigma{\mbox{\protect\boldmath $\sigma$}}
\def\bone{{\mathbf 1}}
\def\vv{{\bf v}}
\def\w{{\bf w}}

\newtheorem{theorem}{Theorem}[section]
\newtheorem{proposition}[theorem]{Proposition}
\newtheorem{lemma}[theorem]{Lemma}
\newtheorem{corollary}[theorem]{Corollary}
\newtheorem{conjecture}[theorem]{Conjecture}


\newenvironment{sarray}{
          \textfont0=\scriptfont0
          \scriptfont0=\scriptscriptfont0
          \textfont1=\scriptfont1
          \scriptfont1=\scriptscriptfont1
          \textfont2=\scriptfont2
          \scriptfont2=\scriptscriptfont2
          \textfont3=\scriptfont3
          \scriptfont3=\scriptscriptfont3
        \renewcommand{\arraystretch}{0.7}
        \begin{array}{l}}{\end{array}}

\newenvironment{scarray}{
          \textfont0=\scriptfont0
          \scriptfont0=\scriptscriptfont0
          \textfont1=\scriptfont1
          \scriptfont1=\scriptscriptfont1
          \textfont2=\scriptfont2
          \scriptfont2=\scriptscriptfont2
          \textfont3=\scriptfont3
          \scriptfont3=\scriptscriptfont3
        \renewcommand{\arraystretch}{0.7}
        \begin{array}{c}}{\end{array}}

\section{Introduction}   \label{sec1}

The antiferromagnetic $q$-state Potts model \cite{Wu_82,Baxter_82,Martin_91} 
has many interesting features. First, for large enough $q$ the model exhibits 
a nonzero ground-state entropy (without frustration). This is important as it
provides an exception to the third law of thermodynamics
\cite{Aizenman_81,Chow_87}. 
Second, the antiferromagnetic Potts model has a rich phase diagram that 
depends explicitly on the lattice structure (in contrast to the universality 
typically enjoyed by ferromagnets). It therefore becomes interesting to
elucidate which features of the lattice structure (e.g., coordination number, 
dimensionality, etc) correlate with the critical properties of the model 
(or the absence of criticality). Furthermore, for each lattice $G$ there is 
a number $q_c(G)$ such that for all $q>q_c$ the Potts model is disordered at 
any temperature (including $T=0$); at $q=q_c(G)$ the system is disordered at 
all positive temperatures and has a zero-temperature critical point. 
Third, the study of Potts antiferromagnets is closely related
to graph theory. Namely, the zero-temperature limit of the antiferromagnetic
Potts model partition function is just the chromatic polynomial $P_G(q)$: 
\be
\lim\limits_{T\rightarrow 0} Z_{G}(q;T) = P_G(q).
\ee
This quantity equals the number of ways of coloring the vertices of the graph
$G$ with $q$ colors, with the constraint that no adjacent vertices have the
same color \cite{Read_88}.
Finally, there are condensed-matter systems that can be 
modeled using Potts antiferromagnets. In particular, the insulator 
SrCr$_{8-x}$Ga$_{4-x}$O$_{19}$ can be described using a three-state 
Kagom\'e-lattice Potts antiferromagnet \cite{Broholm_91,Huse_92}.

One interesting question is the value of the number $q_c(G)$ for the most 
common regular lattices $G$. There are general analytic bounds in terms of the 
coordination number $\Delta$ of the lattice (e.g., $q_c \leq 2\Delta$ 
\cite{Salas_97}), but they are not very sharp. An alternative approach, 
inspired by the Lee--Yang picture of phase transitions \cite{Yang-Lee_52}, is
to study the zeros of the chromatic polynomial when the parameter $q$ is 
allowed to take complex values (see Ref.~\cite{transfer1} for a complete list 
of references). The best results concern families $G_n$ of 
graphs for which the chromatic polynomial can be expressed via a transfer 
matrix of fixed size $M\times M$:
\begin{subeqnarray}
   P_{G_n}(q)  & = &   \tr[ A(q) \, T(q)^n ]  \\[2mm]
               & = &   \sum\limits_{k=1}^{M}
                           \alpha_k(q) \, \lambda_k(q)^n,
\label{general_form_P}
\end{subeqnarray}
where the transfer matrix $T(q)$ and the boundary-condition matrix $A(q)$
are polynomials in $q$, so that the eigenvalues $\{ \lambda_k \}$ of $T$
and the amplitudes $\{ \alpha_k \}$ are algebraic functions of $q$.

The two best understood cases of the Potts
antiferromagnet are the square and triangular lattices.
In this paper we consider the square lattice. For
$q=2$ this model undergoes a finite-temperature second-order phase transition.
For $q=3$ there is strong analytic and numerical support for a zero-temperature
critical point \cite{Lenard_67,Baxter_70,Baxter_82b,Nijs_82,Burton_Henley_97,%
Salas_98,deQueiroz_99,Ferreira_99}, though a rigorous proof is (to our
knowledge) still lacking. For $q=4$ Monte Carlo simulations show that the model
is disordered even at zero temperature \cite{Ferreira_99}.
It thus seems clear that $q_c({\rm sq}) = 3$.

In the transfer matrix approach, the goal is to compute the transfer matrix 
$T$ for a square-lattice strip of width $m$ with some boundary conditions.
With this matrix one can easily compute the chromatic-polynomial zeros of any 
finite strip $m\times n$. In addition, one can also compute the accumulation 
points when $n\rightarrow\infty$. According to the Beraha--Kahane--Weiss 
theorem \cite{BKW_75,BKW_78}, the zeros when $n\rightarrow\infty$ can either 
be isolated limiting points (when the amplitude associated to the dominant
eigenvalue vanishes) or form a limiting curve $\scrb$ (when two dominant
eigenvalues cross in modulus). By studying the limiting curves for different
values of the width $m$ we hope to learn new features of the thermodynamic 
limit $m\rightarrow\infty$.  

The first results for the square lattice using this approach were obtained by 
Shrock and collaborators.\footnote{
  They actually used a generating-function approach which is equivalent to 
  transfer matrices.
}
They studied strips up to $m=6$ (resp.~$m=5$) with 
cylindrical (resp.~free) boundary conditions%
\footnote{Let $m$ (resp.~$n$) denote the number of sites in the transverse
(resp.~longitudinal) direction of the strip, and let F (resp.~P) denote
free (resp.~periodic) boundary conditions in a given direction. Then we
use the terminology:
free         ($m_{\rm F} \times n_{\rm F}$),
cylindrical  ($m_{\rm P} \times n_{\rm F}$),
cyclic       ($m_{\rm F} \times n_{\rm P}$), and 
toroidal     ($m_{\rm P} \times n_{\rm P}$).}
\cite{Shrock_98a,Shrock_98c,Shrock_00d}. They also computed square-lattice
strips with cyclic and toroidal boundary conditions up to $m=4$ 
\cite{Shrock_00d,Shrock_97a,Shrock_99c,Biggs_99b,Shrock_99e,Shrock_00a,%
Shrock_00f}.
The results with free and cylindrical boundary conditions were extended 
to $m\leq 8$ in Ref.~\cite{transfer1}. The qualitative shape of the limiting
curve found for these boundary conditions is quite similar to the one found 
by Baxter in case of the triangular lattice \cite{Baxter_87}, but with the 
zero-temperature critical point lying at $q_c({\rm sq})=3$ rather than
at $q_c({\rm tri})=4$. 

However, there may be an important qualitative difference
between the triangular-lattice and square-lattice cases,
concerning the number of times that the infinite-volume curve $\scrb$
crosses the positive real $q$-axis.  In physical terms this
corresponds to the number of intervals of the real $q$-axis where the
ground-state entropy $S_0(q)$ has a distinct analytic form,
or in other words to the number of ``phases'' that the model has
as $q$ is varied \cite{Shrock_97a,Shrock_97b}.
For many (but not all) finite strip widths $m$,
we find that the limiting curve ${\cal B}$ crosses the 
real axis at a point $q_0(m)$.\footnote{
  For those lattice strips where $q_0$ cannot be strictly defined because
  $\scrb$ does not cross the real axis, the limiting
  curve often includes a pair of complex-conjugate endpoints that are very close
  to the positive real $q$-axis. In these cases, we define $q_0$ to be the
  endpoint closest to that axis with positive imaginary part.
}
It is reasonable to expect that there are infinitely many such widths $m$
and that $q_0(m)$ tends to a limiting value $q_0(\infty)$ as $m \to\infty$.
One could be tempted to assume that $q_0(\infty) = q_c$;
but this, it turns out, is not always true.
Indeed, in the triangular-lattice case, Baxter's exact solution
\cite{Baxter_87} shows that $q_0({\rm tri}) \approx 3.81967$  
while $q_c({\rm tri})=4$; this is because the infinite-volume curve $\scrb$
crosses the real $q$-axis {\em twice}\/ in the interval $0 < q < \infty$.
Thus, in general one cannot identify $q_0$ with $q_c$.\footnote{
  One can qualitatively think of $q_c$ as the limit of the (complex)
  rightmost endpoints of the limiting curve ${\cal B}$ as the lattice strip
  width $m$ increases. These endpoints can be defined in most cases for 
  free boundary conditions in the longitudinal direction 
  \protect\cite{transfer1,transfer3}, but not for periodic boundary 
  conditions in the longitudinal direction \protect\cite{Shrock_BCC}.  
  When $m$ grows large, those endpoints approach the real $q$-axis,
  and in the limit $m=\infty$ they pinch that axis at $q=q_c$.
}
On the other hand, our data for the square lattice suggest that
$q_0(\infty)$ may well be equal to $q_c = 3$.
If this is true, the qualitative shape of the infinite-volume curve $\scrb$
for the square lattice
will be rather different from that found for the triangular lattice.
In particular, region ${\cal D}_2$ in Figure~5 of ref.~\cite{Baxter_87} 
will be ``narrowed'' along the $q$-axis so the points F (i.e., $q_0$) 
and C (i.e., $q_c$) coincide. Thus, region ${\cal D}_2$ will be split into
two complex-conjugate regions that may or may not intersect the positive
real $q$-axis at $q_c$. 
 
In this paper we extend the results of \cite{transfer1} by computing the 
transfer matrices for square-lattice strips of width $9 \le m \le 13$ with 
periodic boundary conditions in the transverse direction and free boundary 
conditions in the longitudinal direction. We have chosen this type of 
boundary conditions because we expect it to provide a faster approach towards
the thermodynamic limit than we would have obtained using free boundary
conditions in both directions,
while at the same time avoiding the technical complications implied by the
imposition of doubly periodic (toroidal) boundary conditions \cite{transfer4}.
The present computation has been possible thanks to a more efficient
algorithm than the one employed in \cite{transfer1};
this improved algorithm allows us to handle
transfer matrices as large as $498\times 498$. The transfer matrices,
of course, enable us to compute the chromatic polynomials
$P_{m \times n}(q)$ for any finite length $n$.
For $9 \le m \le 11$, however, we have been able to obtain the limiting 
curves ${\cal B}$ directly in the limit $n\to\infty$, by using the same
methods as in \cite{transfer1}.
When possible, we have also computed the isolated limiting
zeros for these square-lattice strips and  checked whether the conjectures
put forth in \cite[Section~7]{transfer1} hold true. These conjectures are
related to the Beraha numbers \cite{Beraha_unpub} which seem to play 
a special role in the theory of chromatic polynomials 
(see Ref.~\cite{transfer1} and references therein): 
\be
B_n = 4 \cos^2 {\pi \over n} = 2 + 2 \cos {2\pi \over n}, \ \ \ \
\mbox{for }n=2,3,\ldots.
\label{def_Bn}
\ee

We find that the limiting curves ${\cal B}$ for various system sizes are
qualitatively quite similar; however, they show noticeable differences
depending on whether the width 
$m$ is even or odd. In particular, we conjecture that only the limiting
curves for odd $m$ cross the real axis, thus giving a well-defined value
of $q_0$. Our best estimate for this quantity (assuming monotonicity with the
strip width) reads
\be
q_0({\rm sq}) \gtapprox 2.9161885031
\ee
This estimate is quite close to the expected value for $q_c({\rm sq})=3$. 
For the square-lattice strips with even width $m$ we find for $m\geq 8$ that
there is a small gap separating ${\cal B}$ from the real axis: the real part
of the points closest
to that axis is slightly smaller than the value of $q_0$ for widths $m\pm 1$, 
and based on our numerical data we conjecture that the imaginary part of such
points goes to zero as $B_5^{-m/2}$.

We finally conjecture (based on our numerical evidence) that 
$q_0({\rm sq}) = q_c({\rm sq}) = 3$. This equality has important implications 
on the analytic structure of the free energy in the thermodynamic limit. 
In the zero-temperature triangular-lattice Potts antiferromagnet, we
have that $q_0({\rm tri}) \neq q_c({\rm tri})$ and there are three domains 
of analyticity of the free energy \cite{Baxter_87}. Furthermore, all these
three domains intersect the real $q$-axis; $q_0$ and $q_c$ are the boundaries
of such domains on the positive real $q$-axis. In the zero-temperature
square-lattice Potts antiferromagnet, we have that 
$q_0({\rm sq}) = q_c({\rm sq})$. Thus, on the real $q$-axis there are only 
two domains of definition whose boundary is precisely $q_c$. This is 
a valuable piece of information about the analytic structure of the free
energy of this model whose solution is yet to be found.  

This paper is laid out as follows:
In Section~\ref{sec2} we explain the method used to compute the transfer
matrices. The necessary background can be found in \cite{transfer1}. 
In Section~\ref{sec3} we expose our numerical results
for square-lattice strips with cylindrical boundary conditions.
Finally, Section~\ref{sec4} is dedicated to our conclusions.

\section{Transfer Matrix Algorithm} \label{sec2}

The general theory of Potts model transfer matrices has been
reviewed in \cite[Section~3]{transfer1}, and in what follows we shall
therefore limit ourselves to a rather concise and more informal
description. The main emphasis is on the algorithmic improvements
over Ref.~\cite{transfer1}, which have enabled us to extend the
computations to strips of width $9 \le m \le 13$ with cylindrical
boundary conditions.

Let $G=(V,E)$ be a finite undirected graph with vertex set $V$ and
edge set $E$, and let $\{ J_e \}_{e\in E}$ be a set of coupling constants.
In the Fortuin-Kasteleyn representation \cite{Kasteleyn_69,Fortuin_72}
the partition function of the $q$-state Potts model defined on $G$ reads
\begin{equation}
 Z_G(q,\{v_e\}) = \sum_{E'\subseteq E} q^{k(E')} \prod_{e \in E'} v_e,
\end{equation}
where $k(E')$ is the number of connected components (clusters) in the
subgraph $(V,E')$, and $v_e = \exp(\beta J_e)-1$.
The main advantage of this representation is that $q$ can now be
analytically continued to complex values. However, the price to be paid
is that the Boltzmann weights contain the non-local factor $q^{k(E')}$.

Although this non-locality would seem to inhibit the
construction of the transfer matrix, the problem can be circumvented by
working in a basis of connectivities $\scrc_m$, containing information about
how the $m$ spins in a given time slice $t$ are interconnected in the portion
of $(V,E')$ that is prior to $t$. Clearly, the number $C_m = |\scrc_m|$ of such
connectivities is equal to the number of ways that $m$ points on the rim
of a disc can be connected in the interior of the disc, by means of a
planar graph.
The evaluation of these numbers is a standard combinatorial exercise
\cite{Stanley_86}, and one arrives at the Catalan numbers,
\begin{equation}
 C_m = \frac{(2m)!}{m!(m+1)!}.
\end{equation}

A numerical implementation of the Potts model transfer matrix in the basis
of Catalan connectivities was first described by Bl\"ote and Nightingale
\cite{Blote_82}. These authors also introduced a {\em ranking} of the
connectivities, i.e.~an injective mapping from $\scrc_m$ to
the set of integers $\{1,2,\ldots,C_m\}$. Details on the
construction of the inverse mapping were given in \cite{Jacobsen_98}.
The idea is then to use the integers to index the entries of the transfer
matrix, and the connectivities to work out the possible transitions between
states along with their respective Boltzmann weights. When combined with
a standard sparse matrix factorization, in which the entire transfer matrix
is written as a product of elementary matrices each adding one edge of
$G$, this algorithm is highly efficient, since each non-zero matrix element
can be computed in time $\sim m$.

Since we are considering specifically the zero-temperature antiferromagnet
($v_e=-1$ for all edges $e\in E$) the dimension of the transfer matrix
is actually much less than $C_m$, since among the $m$ points in a given
time slice no pair of neighboring points can belong to the same cluster.
It can be shown \cite[Section~3.3]{transfer1} that the number
$R_m = |\scrr_m|$ of antiferromagnetically allowed states $\scrr_m$
is given by the Riordan numbers, $R_0=1$, $R_1=0$ and
\begin{equation}
 R_m = \sum_{k=0}^{m-1} (-1)^{m-k-1}
       \sum_{j=0}^{\lfloor k/2 \rfloor} {k \choose 2j} C_j
       \ \ \ \mbox{ for } m \ge 2.
\end{equation}
Furthermore, due to isotropy and the periodic boundary conditions, the
connectivities $\scrr_m$ are invariant under the action of the dihedral group
$D_m$ on the cyclically ordered set of $m$ points in the time
slice considered. The appropriate basis states $\scrs_m$ are thus found
from the Riordan states $\scrr_m$
by symmetrizing with respect to the elements of $D_m$.
The resulting dimension of the transfer matrix
$\mbox{SqCyl}($m$) = |\scrs_m|$ has been
evaluated in \cite[Table~2]{transfer1}.

However, the sparse matrix decomposition does not respect any of these
constraints, and accordingly the number of connectivities in the intermediary
states between to subsequent time slices will still be $C_m$.
To compute a given column $T(q) |s_{\rm i}\rangle$ of the transfer matrix,
with $|s_{\rm i}\rangle \in \scrs_m$, we therefore proceed as follows.
First, each of the distinct states in the set
$\{ D_m^{(j)} |s_{\rm i}\rangle : j=1,2,\ldots,2m\} \subseteq \{C_m\}$
is assigned
a weight $q^{k(|s_{\rm i}\rangle)}$, where $k(|s_{\rm i}\rangle)$ is the
number of connected components in $|s_{\rm i}\rangle$.
Second, we act on this initial state with $T(q)$,
or rather with its sparse matrix decomposition.
The final state obtained in this manner should again be symmetric with
respect to all $D_m^{(j)}$, and it provides a non-trivial check of our
algorithm
that this is actually the case. Reading now off the weight of each state
$\langle s_{\rm f}| \in \scrs_m$, and multiplying by
$q^{-k(\langle s_{\rm f}|)}$,
we deduce the value of the matrix element
$\langle s_{\rm f} | T(q) | s_{\rm i} \rangle$.

A few practical remarks are in order.
Since all weights are polynomials in $q$ the computations are
done using symbolic algebra, representing the integer coefficients
of each power of $q$ in an array. These integers may in principle be very
large, and thus cause overflow in a standard 32 bit integer arithmetic.
This problem
can be easily remedied by using modular arithmetic, i.e.~computing all
coefficients modulo different primes, and retrieving the complete result
from the Chinese remainder theorem. However, for the system sizes used
in this work it turned out that 32 bits are actually sufficient.

As usually with transfer matrix methods, the main obstacle for going to
large system sizes is due to memory limitations. Our algorithm for $m=13$
used some 500 MBytes of memory, and although one might have gone to
slightly larger systems on a bigger computer, we refrained from doing so
because of the restrictions imposed by the subsequent analysis
(see Section~\ref{sec3}).
The efficiency of our algorithms can be appreciated by noting that the
calculations for $m=8$, the largest system size treated in \cite{transfer1},
were done in less than ten seconds.
In fact we have reproduced all results given in Ref.~\cite{transfer1}
for the transfer matrices and the boundary condition vectors (see below)
with $m \le 8$; besides validating Ref.~\cite{transfer1}, this is of
course also a nontrivial check on the correctness of our programs.

Apart from the transfer matrix we are also interested in computing the
zero-temperature partition functions (i.e., the chromatic polynomials) 
$P(m_{\rm P} \times n_{\rm F})$ for strips of a finite
length $n$ and with free boundary conditions in the longitudinal
direction. As explained in \cite[Section~6]{transfer1} this can be
achieved by sandwiching powers of $T(q)$ between suitable initial and final
vectors,
\begin{equation}
 P(m_{\rm P} \times n_{\rm F}) =
 \langle v_{\rm f} | T(q)^{n-1} | v_{\rm i} \rangle.
\end{equation}

The initial vector $|v_{\rm i}\rangle$ is simply the unit vector defined
by assigning a weight one to the unique state in which each of the $m$
points belongs to a distinct cluster, and zero to all other states.
The final vector $\langle v_{\rm f}|$ corresponds to adding the last row of
the lattice, followed by a trace over the final states $\langle s_{\rm f} |$
in which each state is weighted by $q^{k(\langle s_{\rm f} |)}$.
It is found by a procedure very similar
to the one used to compute the matrix elements
$\langle s_{\rm f} | T(q) | s_{\rm i} \rangle$.
The only difference is that, in the last step, we take the trace over
the $\langle s_{\rm f}|$, rather than projecting on each of them and
multiplying by $q^{-k(\langle s_{\rm f}|)}$.

\section{Numerical Results for the Square-Lattice Chromatic Polynomial:
   \hfill\break Periodic Transverse Boundary Conditions}
   \label{sec3}

We have computed the transfer matrix $T(q)$ and the limiting curves $\scrb$
for square-lattice strips of widths $9 \leq L_x \leq 13$
with periodic boundary conditions in the transverse direction. One direct check
of our results is provided by the (trivial) identity
\be
Z(m_{\rm P}\times n_{\rm F}) = Z(n_{\rm F} \times m_{\rm P}) 
\ee
The chromatic polynomials $Z(n_{\rm F} \times m_{\rm P})$ with $n=2,3$ 
are well-known \cite{BDS,Shrock_00a}. We have 
used both the resultant and the direct-search methods to obtain the limiting
curves (see \cite{transfer1} for details). To compute the isolated limiting 
points we define the matrix $D(q)$:
\be
   D(q)  \;=\;
        \left( \begin{array}{cccc}
                    P_{n\times 1} & P_{n\times2} & \cdots & P_{n\times M} \\
                    P_{n\times2}  & P_{n\times3} & \cdots & P_{n\times M+1}\\
                    \vdots & \vdots &        & \vdots   \\
                    P_{n\times M} & P_{n\times M+1}& \cdots & P_{n\times 2M-1} 
               \end{array}
        \right),   
\label{def_Dq}
\ee
where $M$ is the dimension of the corresponding transfer matrix. Then it 
can be proved \cite{Beraha_80} that 
\be
\det D(q) = \prod\limits_{k=1}^M \alpha_k \prod\limits_{1\leq i < j \leq M}
   (\lambda_j - \lambda_i)^2.
\label{def_detD}
\ee
Thus, all the zeros of the amplitudes $\alpha_k$ are also zeros of 
$\det D(q)$ (see also \cite[Section 2.2]{transfer1}).

%
%
\subsection{$L_x = 9_{\rm P}$} \label{sec9P}

The transfer matrix is 22-dimensional;
it can be found in the {\sc Mathematica} file {\tt transfer2.m} 
that is available with the electronic version of this paper in the
{\tt cond-mat} archive at
{\tt http://www.lanl.gov}.
We have checked that none of the amplitudes vanishes identically. 

\begin{figure}
  \centering
  \epsfxsize=380pt\epsffile{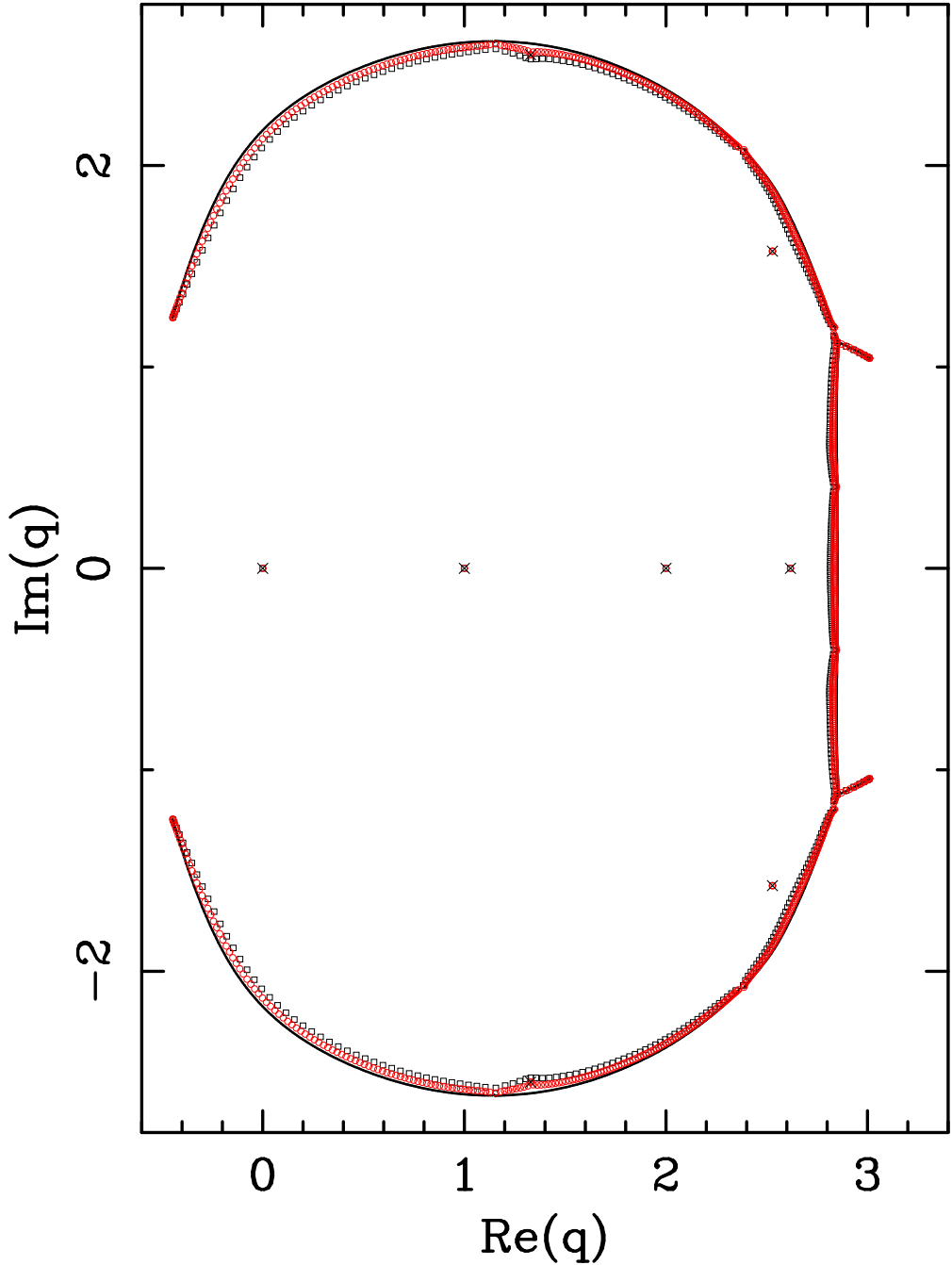}
  \caption[a]{\protect\label{Figure_sq_9PxInftyF}
  Zeros of the partition function of the $q$-state Potts antiferromagnet
  on the square lattices $9_P \times 45_F$ (squares),
  $9_P \times 90_F$ (circles) and $9_P\times\infty_F$ (solid line).
  The isolated limiting zeros are depicted by a $\times$.
  The limiting curve was computed using the direct-search method, except 
  the endpoints that were computed using the resultant method.
  }
\end{figure}

We have computed the limiting curve ${\cal B}$ using the direct-search 
method. In addition, we were able to compute the resultant at $\theta=0$
(but unfortunately not for all $\theta$).
Some of its zeros are precisely the endpoints of the limiting curve, which is 
why our results for the endpoints are more accurate in this section than 
in the following ones.\footnote{ 
   The zeros of the resultant at $\theta=0$ correspond to values of $q$ 
   such that there are two (or more) equal eigenvalues. However, we are
   only interested in the values of $q$ associated to {\em dominant} 
   equal eigenvalues.  
}  
The limiting curve contains nine disconnected pieces (see 
Figure~\ref{Figure_sq_9PxInftyF}). One of them crosses   
the real axis at $q\approx 2.8505722621$. There are 20 endpoints: 
\begin{subeqnarray}
q &\approx&           -0.44699181 \pm 1.24252295\,i \\
\slabel{endpoint_9P_2}
q &\approx& \phantom{-}1.14912991 \pm 2.61264326\,i\, ;  \quad  
q  \approx             1.15312070 \pm 2.61981372\,i \\
\slabel{endpoint_9P_3}
q &\approx& \phantom{-}2.36507014 \pm 2.08719033\,i\, ; \quad  
q  \approx             2.38763742 \pm 2.08307229\,i \\
\slabel{endpoint_9P_4}
q &\approx& \phantom{-}2.83939211 \pm 1.19488272\,i\, ; \quad 
q  \approx             2.83111413 \pm 1.16708976\,i \\
\slabel{endpoint_9P_5}
q &\approx& \phantom{-}2.84682005 \pm 0.40527862\,i\, ; \quad 
q  \approx             2.85144522 \pm 0.40515929\,i \\
q &\approx& \phantom{-}3.01123194 \pm 1.04259193\,i
\end{subeqnarray}
%
There are eight small gaps defined by the endpoints given in 
Eqs.~(\ref{endpoint_9P_2}--\ref{endpoint_9P_5}).  
Finally, there are two T-points at $q\approx 2.854 \pm 1.123\,i$. 

We have also computed the polynomial $\det D(q)$ for this strip. It can be 
written as
\begin{eqnarray}
\det D(q) &=& q^{22} (q-1)^{52} (q-2)^{22} (q^2-3q+1)^{19} (q-3)^{42} 
            (q^3 - 5q^2 + 6q -1)^7 \nonumber \\ 
          & & \qquad (q^2 - 4q + 2)^4 (q^3 - 6q^2 + 9q -1)
                     (q^2 - 5q + 5) P(q)^2  ,
\label{detD_9P}
\end{eqnarray}
where $P(q)$ is a polynomial of degree 1170 with integer coefficients (and 
not factorizable over the integers). The
full expression of \reff{detD_9P} is included in the {\sc Mathematica} file
{\tt transfer2.m}. The pre-factors appearing in \reff{detD_9P} correspond to 
the first nine minimal polynomials $p_n(q)$ given in
\cite[Table 1]{transfer1}. In particular, this means that
$\det D(q)=0$ for the first nine Beraha numbers 
$B_2,\ldots,B_{10}$,
in agreement with Conjecture~7.1 of \cite{transfer1}. 
Indeed, the first three Beraha numbers $0,1,2$ are trivial zeros as all the 
amplitudes vanish at those values. Thus, they are isolated limiting points.
The non-trivial isolated limiting points are the Beraha number 
$B_5=(3+\sqrt{5})/2$, and the complex conjugate pairs 
$q\approx 1.3290624037 \pm 2.5498775371 \,i$ and 
$q\approx 2.5285781385 \pm 1.5750286926 \,i$ (see 
Figure~\ref{Figure_sq_9PxInftyF}).

The convergence to the non-trivial real zero $B_5$ is quite fast (actually, 
exponential) as shown in Table~\ref{table_zeros_cyl}. The largest real 
zero in Table~\ref{table_zeros_cyl} converges at an approximate $1/n$ rate 
to the value $q_0 \approx 2.8505722621$. 

We have also checked Conjectures 7.2 and 7.3 of \cite{transfer1}, namely that
$\det D(q) > 0$ for all $q=B_k$ with $k>10$. In particular, we have checked 
this result up to $k=50$. We expect that the same result holds also
for $k>50$ as
$f(k) = \det D(q=B_k)$ is clearly a increasing monotonic function  
of $k$. 

%
%
\subsection{$L_x = 10_{\rm P}$} \label{sec10P}

The transfer matrix is 51-dimensional;
it can be found in the {\sc Mathematica} file {\tt transfer2.m}.
We have checked that none of the amplitudes vanishes identically. 
Unfortunately we were unable to compute the resultant even at $\theta=0$,
so we used a pure direct-search method.

The limiting curve ${\cal B}$ contains ten connected 
pieces. None of them  crosses the real axis. The points closest to that axis 
are  $q\approx 2.82813 \pm 0.00097\,i$. 
We find the following 22 approximate endpoints:
\begin{subeqnarray}
q &\approx&           -0.48079 \pm 1.13785\,i \\        
\slabel{endpoint_10P_2}
q &\approx& \phantom{-}0.94694 \pm 2.61250\,i \,; \quad 
q  \approx  \phantom{-}0.94561 \pm 2.60984\,i \\        
\slabel{endpoint_10P_3}
q &\approx& \phantom{-}2.18917 \pm 2.26882\,i \,; \quad 
q  \approx  \phantom{-}2.19871 \pm 2.26845\,i \\        
\slabel{endpoint_10P_4}
q &\approx& \phantom{-}2.76451 \pm 1.49846\,i \,; \quad 
q  \approx  \phantom{-}2.77167 \pm 1.48200\,i \\        
\slabel{endpoint_10P_5}
q &\approx& \phantom{-}2.82813 \pm 0.00097\,i \\        
\slabel{endpoint_10P_6}
q &\approx& \phantom{-}2.86581 \pm 0.87778\,i \,; \quad 
q  \approx  \phantom{-}2.87243 \pm 0.87284\,i \\        
q &\approx& \phantom{-}3.21691 \pm 0.52738\,i           
\end{subeqnarray} 
%
There are nine small gaps defined by the endpoints 
(\ref{endpoint_10P_2}--\ref{endpoint_10P_6}).
Finally, there is a T-point close to $q\approx 2.874 \pm 0.941\,i$. 

\begin{figure}
  \centering
  \epsfxsize=380pt\epsffile{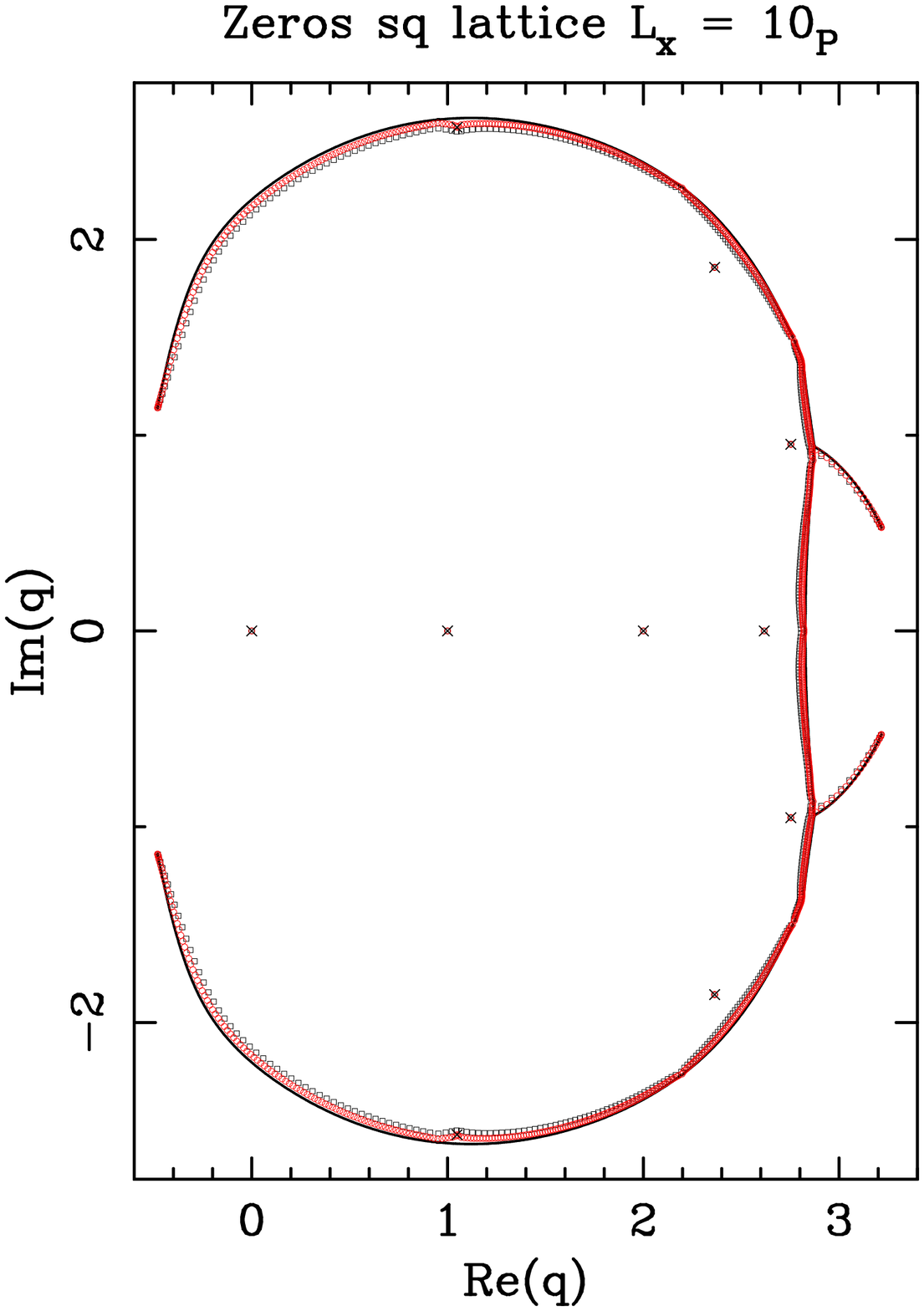}
  \caption[a]{\protect\label{Figure_sq_10PxInftyF}
  Zeros of the partition function of the $q$-state Potts antiferromagnet
  on the square lattices $10_P \times 50_F$ (squares),
  $10_P \times 100_F$ (circles) and $10_P\times\infty_F$ (solid line).
  The isolated limiting zeros are depicted by a $\times$.
  The limiting curve was computed using the direct-search method. 
  }
\end{figure}

In this case we have been unable to obtain the polynomial $\det D(q)$ due
to lack of computer memory.   
By inspection of Figure~\ref{Figure_sq_10PxInftyF} we expect four real
limiting points at the first four Behara numbers $q=0,1,2,B_5$. 
The first two limiting points (i.e., $q=0,1$) are trivial ones, as all the 
amplitudes vanish. We have computed the amplitudes $\alpha_k$ corresponding
to $q=2,B_5$ and, in both cases, the dominant amplitude $\alpha_\star$ 
vanishes.  
The convergence to the non-trivial real zeros $q=2,B_5$ 
is quite fast as shown in Table~\ref{table_zeros_cyl}.
There are also three pairs of complex conjugate isolated limiting points: 
$q \approx 2.7535938496 \pm 0.9533568270\,i$, 
$q \approx 2.3640435205 \pm 1.8567782883\,i$, and  
$q \approx 1.0488718011 \pm 2.5727097899\,i$.  
We have computed the amplitudes corresponding to these values of $q$. 
The dominant amplitude is very small for the first two pairs (the absolute
value of that amplitude is respectively $\approx 4.09 \times 10^{-32}$, and
$\approx 6.4 \times 10^{-42}$). The dominant amplitude for the last pair 
is not that small ($|\alpha_\star| \approx 3.03 \times 10^{-4}$). 
We believe there is an 
isolated limiting point nearby, but we cannot guess a better estimate as 
the convergence is still dominated by the $1/n$ ratio coming from the regular
points of ${\cal B}$. This phenomenon also occurs for $L_x=9_{\rm P}$ (see 
the upper isolated limiting point in Figure~\ref{Figure_sq_9PxInftyF}). 

We have also computed the amplitudes $\alpha_k$ at the first Beraha numbers
$B_n$ with $n\leq 22$. For $B_2,\ldots,B_5$ the dominant amplitude
vanishes as explained above. For $B_6,\ldots,B_{11}$ we find at least one
vanishing amplitude, but none of them is the dominant one. Thus, none of these
Beraha numbers is an isolated limiting point. Finally, for 
$B_{12},\ldots,B_{22}$ all the amplitudes are nonzero. Furthermore, the 
product of all the amplitudes for these Beraha numbers is positive; this 
implies that $\det D(q) > 0$ for $q=B_{12},\ldots,B_{22}$ because of  
Eq.~\reff{def_detD}. These results support Conjectures 7.1, 7.2, and 7.3 of 
\cite{transfer1}.  

%
%
\subsection{$L_x = 11_{\rm P}$} \label{sec11P}

The transfer matrix is 95-dimensional;
it can be found in the {\sc Mathematica} file {\tt transfer2.m}.
We have checked that none of the amplitudes vanishes identically.

The limiting curve crosses the real axis at $q\approx 2.8900930977$. 
We find the following 20 approximate endpoints: 
\begin{subeqnarray}
q &\approx&           -0.50104 \pm 1.05084\,i \\        
\slabel{endpoint_11P_2}
q &\approx& \phantom{-}0.77281 \pm 2.58699\,i \,; \quad 
q  \approx  \phantom{-}0.77329 \pm 2.58795\,i \\        
\slabel{endpoint_11P_3}
q &\approx& \phantom{-}2.01541 \pm 2.39925\,i \,; \quad 
q  \approx  \phantom{-}2.01168 \pm 2.39892\,i \\        
\slabel{endpoint_11P_4}
q &\approx& \phantom{-}2.66452 \pm 1.73992\,i \,; \quad 
q  \approx  \phantom{-}2.67029 \pm 1.73369\,i \\        
\slabel{endpoint_11P_5}
q &\approx& \phantom{-}2.88992 \pm 0.35265\,i \,; \quad 
q  \approx  \phantom{-}2.89068 \pm 0.35252\,i \\        
q &\approx& \phantom{-}3.09853 \pm 0.85628\,i           
\end{subeqnarray}
%
The endpoints (\ref{endpoint_11P_2}--\ref{endpoint_11P_5}) define eight
small gaps. 
Finally, there is a pair of complex conjugate T-points close to
$q\approx 2.902 \pm 1.015\,i$. 

\begin{figure}
  \centering
  \epsfxsize=380pt\epsffile{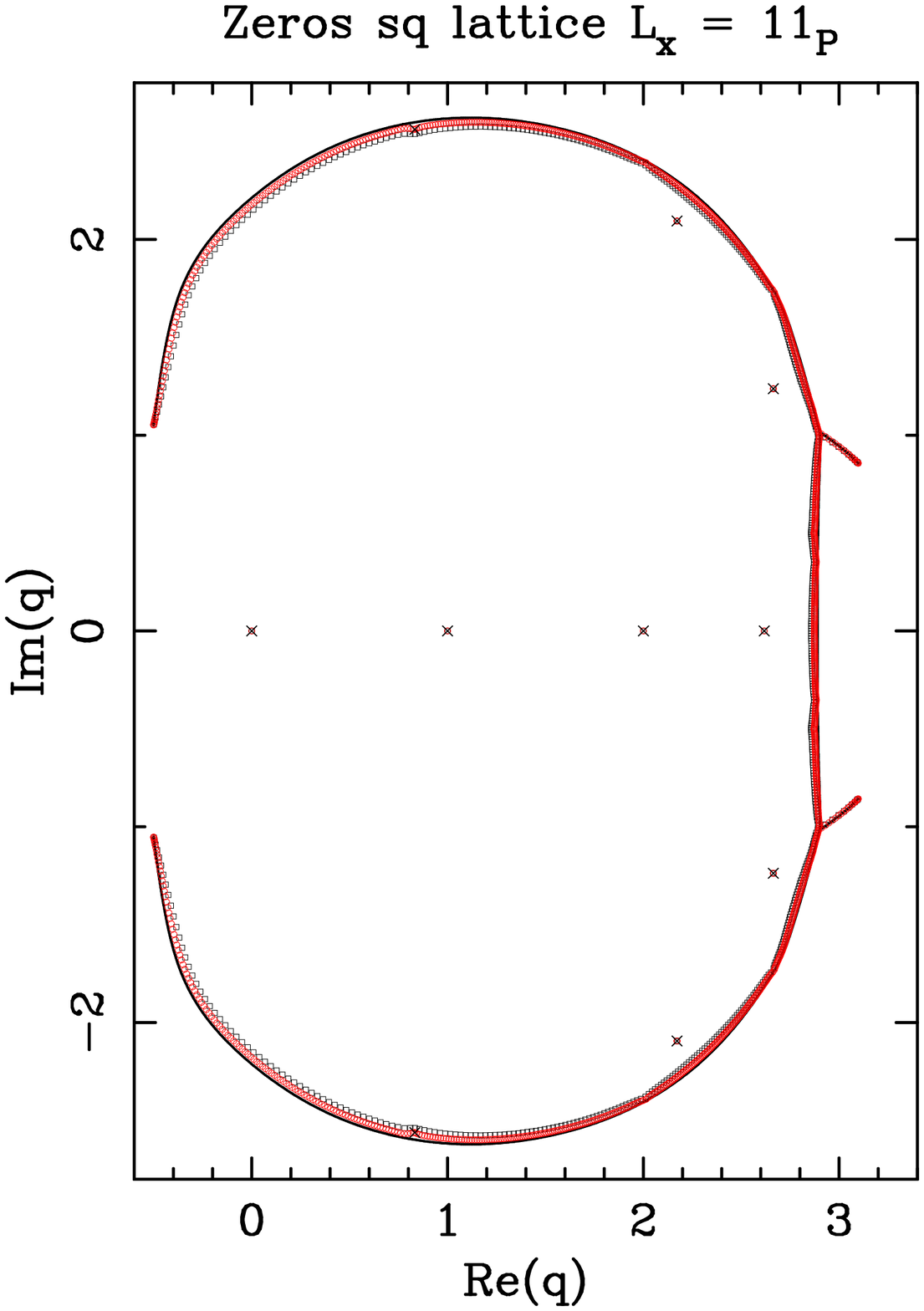}
  \caption[a]{\protect\label{Figure_sq_11PxInftyF}
  Zeros of the partition function of the $q$-state Potts antiferromagnet
  on the square lattices $11_P \times 55_F$ (squares),
  $11_P \times 110_F$ (circles) and $11_P\times\infty_F$ (solid line).
  The isolated limiting zeros are depicted by a $\times$.
  The limiting curve was computed using the direct-search method. 
  }
\end{figure}

By inspection of Figure~\ref{Figure_sq_11PxInftyF} we expect four real
limiting points at the first four Beraha numbers $q=0,1,2,B_5$. 
The first three values (i.e., $q=0,1,2$) correspond to trivial isolated 
limiting points as all the amplitudes vanish. The fourth value $q=B_5$ has
a vanishing dominant amplitude; thus, it is a true isolated limiting point. 
The convergence to $q=B_5$ is quite fast as shown in 
Table~\ref{table_zeros_cyl}. The largest real
zero in Table~\ref{table_zeros_cyl} converges at an approximate $1/n$ rate
to the value $q_0 \approx 2.8900930977$.
There are also three pairs of complex conjugate isolated limiting points: 
$q \approx 2.6642077467 \pm 1.2375206436\,i$, 
$q \approx 2.1718870063 \pm 2.0937620001\,i$, and  
$q \approx 0.8331263199 \pm 2.5614743076\,i$.  
The value of the dominant amplitude vanishes (within our numerical precision)
for the first two pairs (namely, $|\alpha_\star| \sim 10^{-30}$). However,
the last pair is close, but not quite equal, to the real isolated limiting
point. In this case we have $|\alpha_\star| \approx 4.70 \times 10^{-3}$.
We believe that there is an isolated limiting point nearby, but that
the convergence to 
this point is still dominated by the $1/n$ rate coming from the nearby regular
points in ${\cal B}$. 

We have also checked that Conjectures 7.1, 7.2, and 7.3 of \cite{transfer1} 
hold for this strip: there is at least one vanishing amplitude for
$q=B_2,\ldots,B_{12}$, while all the amplitudes are nonzero for
$q=B_{13},\ldots,B_{22}$. Only for $q=B_2,\ldots,B_5$ the dominant amplitude
$\alpha_\star$ vanishes; thus, these are the only real isolated limiting
points. Finally, the product $\prod_k \alpha_k > 0$
for $q=B_{13},\ldots,B_{22}$, implying that $\det D(q)>0$. 

%
%
\subsection{$L_x = 12_{\rm P}$} \label{sec12P}

\begin{figure}
  \centering
  \epsfxsize=380pt\epsffile{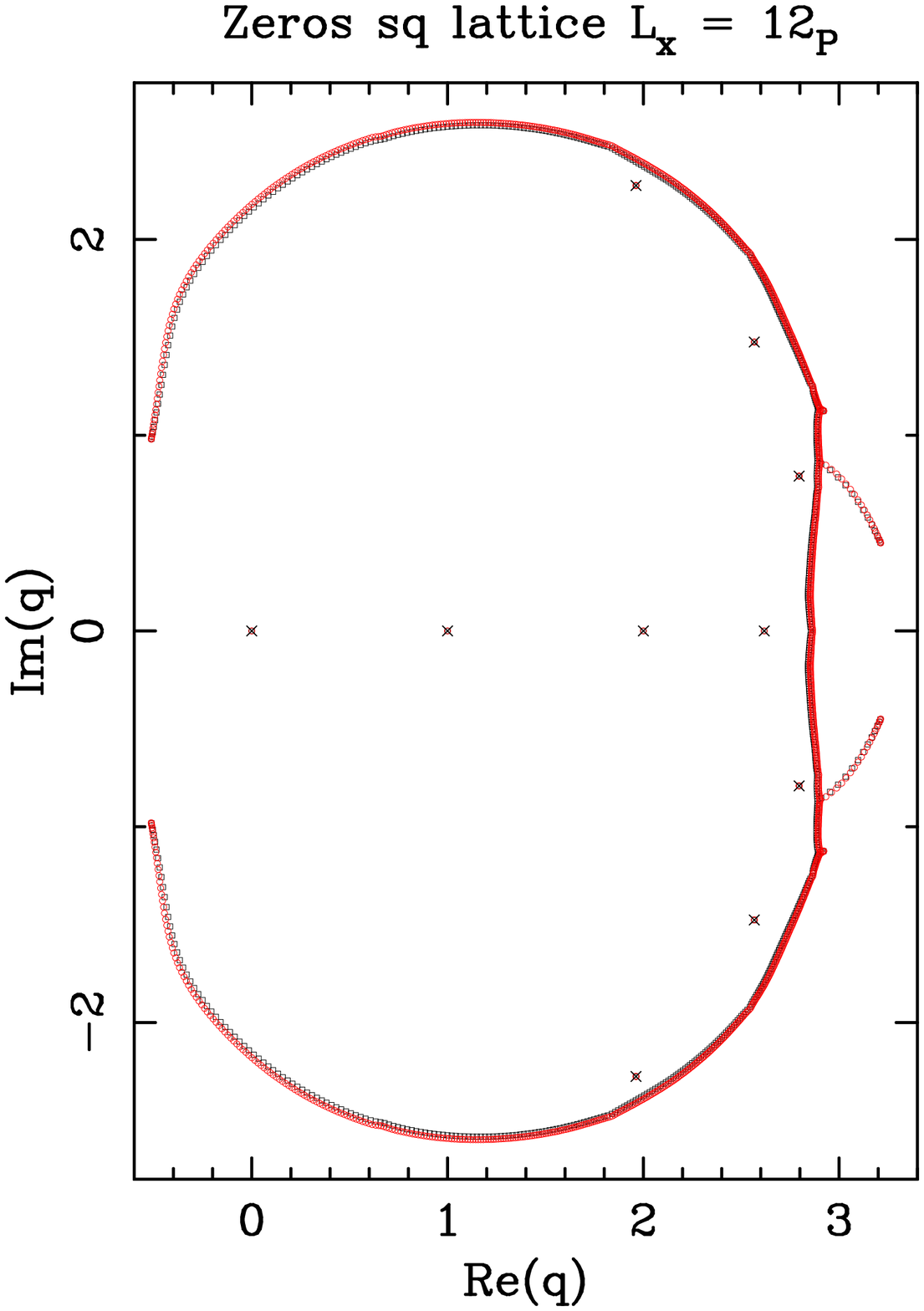}
  \caption[a]{\protect\label{Figure_sq_12PxInftyF}
  Zeros of the partition function of the $q$-state Potts antiferromagnet
  on the square lattices $12_P \times 60_F$ (squares) and 
  $12_P \times 96_F$ (circles).  
  The isolated limiting zeros are depicted by a $\times$.
  }
\end{figure}

The transfer matrix is 232-dimensional;
it can be found in the {\sc Mathematica} file {\tt transfer2.m}. We have
checked that none of the amplitudes vanish identically. 
We have not been able to compute the limiting curve ${\cal B}$ for this case 
as the transfer matrix is too large. However, we could obtain the chromatic 
polynomials for square-lattice strips with aspect ratios between 1 and 8 
(i.e., $P_{12\times12}$, $P_{12\times24}$, \ldots, $P_{12\times96}$). As
shown in Figure~\ref{Figure_sq_12PxInftyF}, the zeros converge to a 
limiting curve very similar to the ones obtained in the previous cases.   
Via the direct-search method (applied to the limiting curve) we have been 
able to obtain an estimate of the
point of the limiting curve ${\cal B}$ which is closest to the real axis,
reading $q\approx 2.874356 \pm 0.00037\,i$. 

By inspection of Figure~\ref{Figure_sq_12PxInftyF} we conclude that
there are four real isolated limiting points at the first four Beraha numbers. 
The first two of those points (i.e., $q=0,1$) are trivial as they correspond
to the vanishing of all the amplitudes. The convergence to the other two 
(non-trivial) real limiting points is quite fast as can be observed from
Table~\ref{table_zeros_cyl}. We also find that the dominant amplitudes for 
$q=2,B_5$ are very small. This evidence supports our belief that 
$q=2$ and $B_5$ are true isolated limiting points.
We also find by inspection three pairs of complex conjugate
isolated limiting points:  
$q\approx 1.9629324150 \pm 2.2751179548\,i$, 
$q\approx 2.5667183607 \pm 1.4761470034\,i$, and
$q\approx 2.7956875974 \pm 0.7913632912\,i$.
The values of $|\alpha_\star|$ are very small for these values of $q$: 
$1.11\times 10^{-15}$, $5.29\times 10^{-20}$, and $1.22\times 10^{-16}$ 
respectively.

We have also checked that Conjectures 7.1, 7.2, and 7.3 of \cite{transfer1} 
hold for this strip: there is at least one vanishing amplitude for 
$q=B_2,\ldots,B_{13}$, whereas all the amplitudes are nonzero for 
$q=B_{14},\ldots,B_{22}$. Furthermore, for these latter values 
$\prod_k \alpha_k >0$, thus $\det D(q) > 0$ \reff{def_detD}.  

%
%
\subsection{$L_x = 13_{\rm P}$} \label{sec13P}

\begin{figure}
  \centering
  \epsfxsize=380pt\epsffile{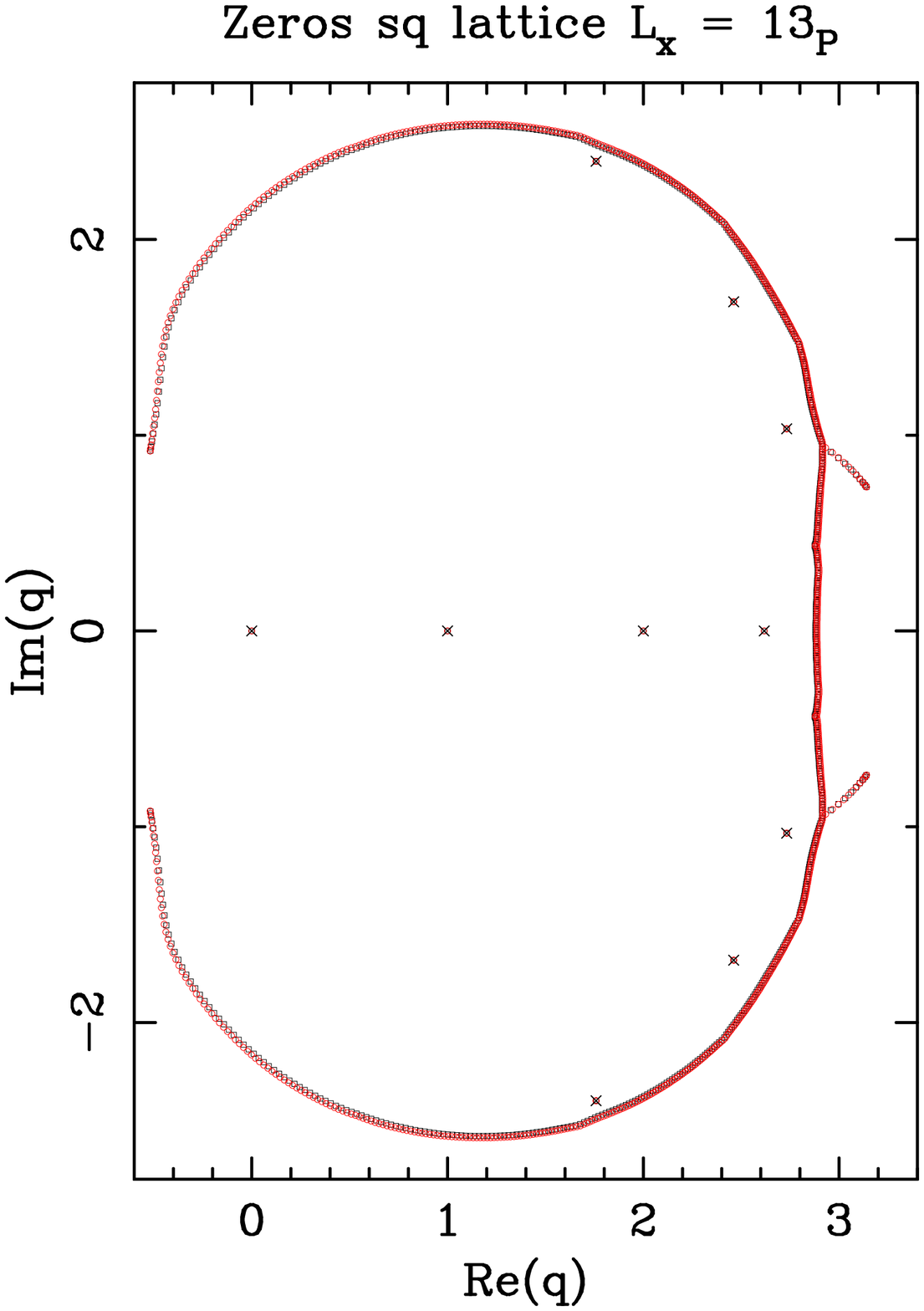}
  \caption[a]{\protect\label{Figure_sq_13PxInftyF}
  Zeros of the partition function of the $q$-state Potts antiferromagnet
  on the square lattices $13_P \times 52_F$ (squares) and
  $13_P \times 65_F$ (circles).
  The isolated limiting zeros are depicted by a $\times$.
  }
\end{figure}

The transfer matrix is 498-dimensional;
it can be found in the {\sc Mathematica} file {\tt transfer2.m}.
We have checked that none of the amplitudes vanish identically.
We have not been able to compute the limiting curve ${\cal B}$ for this case
as the transfer matrix is too large. However, we could obtain the chromatic
polynomials for square-lattice strips with aspect ratios between 1 and 5
(i.e., $P_{13\times13}$, \ldots, $P_{13\times65}$). As
shown in Figure~\ref{Figure_sq_13PxInftyF}, the zeros converge to a
limiting curve very similar to the ones obtained in the previous cases.
Even though the computation of the full limiting curve is too time consuming,
it is very interesting to have an estimate for the value where 
${\cal B}$ crosses the real axis. Our result is $q_0 \approx 2.9161885031$,
using the direct search method.  

By inspection of Figure~\ref{Figure_sq_13PxInftyF} we conjecture that
there are 4 real isolated limiting points corresponding to the first four
Beraha numbers $q=0,1,2,B_5$. The first three are trivial zeros as all the
amplitudes vanish. From Table~\ref{table_zeros_cyl} we observe that the 
convergence to $q=B_5$ is very fast. We have also computed the 
dominant amplitude for $q=B_5$. This indeed vanishes, thus $q=B_5$ is a 
true isolated limiting point. 
The largest real zero in Table~\ref{table_zeros_cyl} converges at an 
approximate $1/n$ rate to the value $q_0 \approx 2.9161885031$.
We also find empirically three pairs
of complex conjugate zeros that are likely to be isolated limiting points.
As the convergence to those points is exponentially fast, the estimates
obtained from the chromatic zeros for the lattice $13_{\rm P}\times 53_{\rm F}$
are expected to be close enough to the true values: 
$q\approx 1.7591342818 \pm 2.3988734769\,i$, 
$q\approx 2.4614147948 \pm 1.6813861639\,i$, and
$q\approx 2.7323074066 \pm 1.0328078058\,i$. Indeed, we obtain very small 
dominant amplitudes for all three values of $q$, namely 
$|\alpha_\star| \approx 7.48\times 10^{-14}$,
$1.30\times 10^{-23}$, and $1.18\times 10^{-21}$ respectively.

We have also checked that Conjectures 7.1, 7.2, and 7.3 of \cite{transfer1}
hold for this strip: there is at least one vanishing amplitude for
$q=B_2,\ldots,B_{14}$, whereas all the amplitudes are nonzero for
$q=B_{15},\ldots,B_{22}$. Furthermore, for these latter values
$\prod_k \alpha_k >0$, thus $\det D(q) > 0$ \reff{def_detD}.

\section{Conclusions} \label{sec4}

\begin{figure}
  \centering
  \epsfxsize=380pt\epsffile{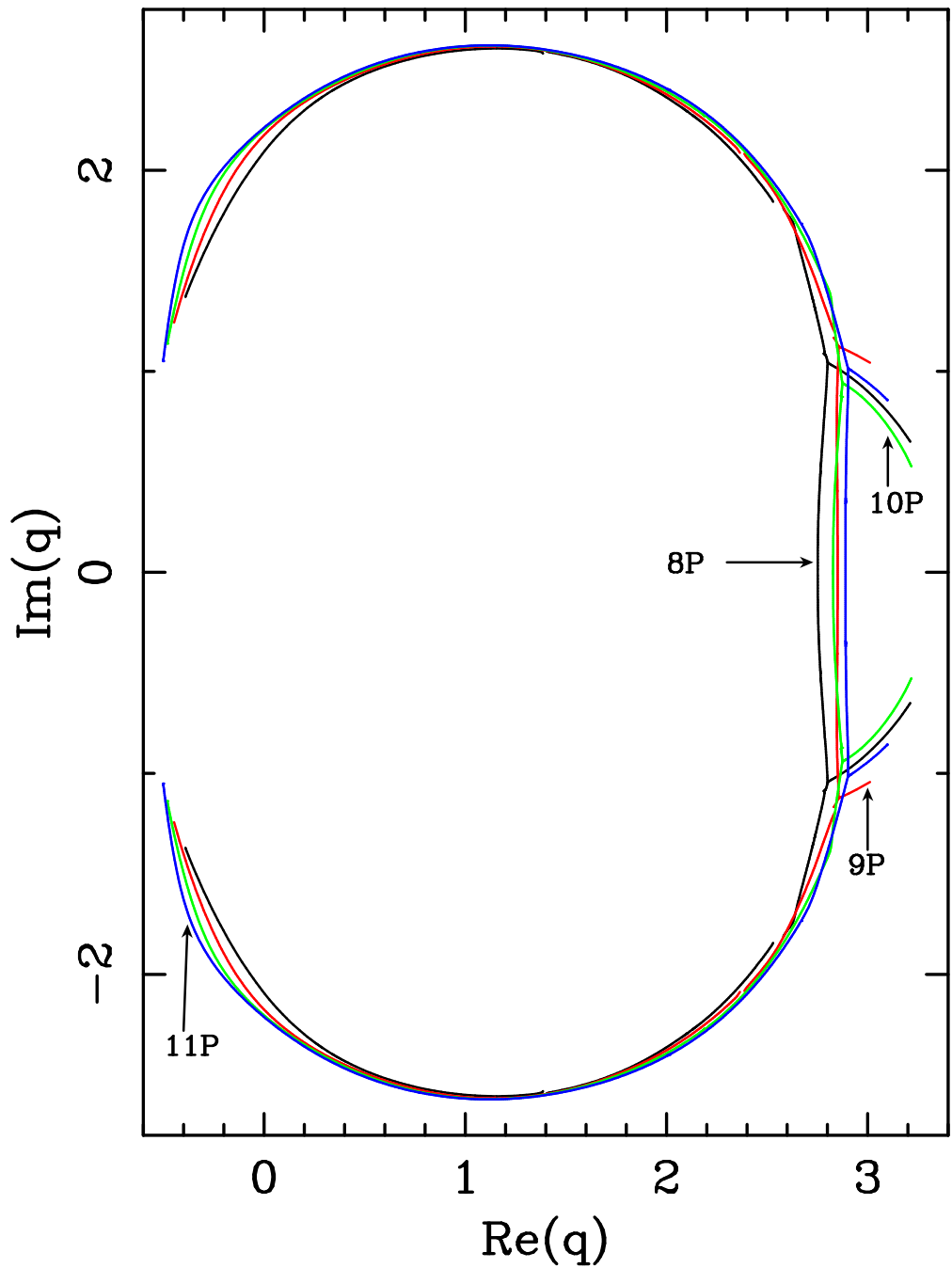}
  \caption[a]{\protect\label{Figure_sq_all}
  Limiting curves for the square-lattice strips 
  $L_{\rm P} \times \infty_{\rm F}$ with $8 \leq L \leq 11$. The curve for
  $L=8$ was obtained in \protect\cite{transfer1}.
  }
\end{figure}

We have computed the transfer matrix of the zero-temperature Potts
antiferromagnet defined on square-lattice strips of widths
$9 \leq m \leq 13$ with cylindrical boundary conditions;
the corresponding result with free boundary conditions are available
from the authors upon request.
For $m=9,10,11$ we have been able to compute the 
limiting curves of the partition function zeros
using the direct-search method. All of them have a similar
qualitative shape (see Figure~\ref{Figure_sq_all}) although we observe 
slight differences between the limiting curves with even width and those with
odd width. 

In Table~\ref{table_summary} we summarize the basic features of
the limiting curves. It is worth noting that the number of connected 
components and endpoints increases with the strip width $m$. 
However, the gaps between those endpoints narrow as $m$ increases.  
Thus, the thermodynamic limit $m\rightarrow\infty$ seems to be rather 
complicated in this model. This may be the cause the solution to this model
is still lacking. On the other hand, the thermodynamic limit for the 
triangular-lattice case is smoother \cite{transfer3} and the model is solvable 
\cite{Baxter_87}.

The differences between even- and odd-width strips is manifest when we 
try to define $q_0$.  
Only the curves corresponding to odd width cross the real 
axis (see Table~\ref{table_summary}). Thus, the value of $q_0$ can be defined
for those strips. We also observe that the value of $q_0$ is monotonically 
increasing with the width $m$. Thus we conjecture that

\begin{conjecture}
\label{conj_odd1} 
The limiting curves for square-lattice strips with cylindrical boundary 
conditions and odd width $m$ cross the 
real $q$-axis at a point $q_0(m)$. The function $q_0(m)$ is a monotonically 
increasing function of $m$.
\end{conjecture} 

Our best estimate for $q_0$ (assuming that Conjecture~\ref{conj_odd1} holds) is 
\be
q_0({\rm sq}) \gtapprox 2.9161885031 \, .
\label{best_q0}
\ee
This estimate comes from the strip with $L_x=13_{\rm P}$. This value of $q_0$
is surprisingly close to the expected value for $q_c({\rm sq})=3$
for this lattice. (It is closer to $q_c$ than in case of the triangular
lattice where $q_0({\rm tri}) \approx 3.81967$ and $q_c({\rm tri})=4$.)
It might happen that for the square lattice $q_0=q_c$. If this were
the case, the limiting curve ${\cal B}$ when $m\rightarrow\infty$ would
be qualitatively different from the triangular-lattice limiting curve
computed by Baxter \cite{Baxter_87}. In particular, instead of the three
domains of definition of the free energy on the positive real $q$-axis for the
triangular-lattice case, we would have only two such domains.  

In Table~\ref{table_summary} we also observe that for even strip widths $m$
the limiting curve does not cross the real axis. We can nevertheless define
$q_0$ to be the closest endpoint to the real $q$-axis (with positive imaginary
part). The real part of this quantity seems to be also a monotonically 
increasing 
function of the lattice width, while the imaginary part goes rather quickly 
to zero. Assuming the monotonicity of $\real q_0$ our best estimate for
$q_0$ would be $q_0({\rm sq}) \gtapprox 2.87436$ which is smaller than 
\reff{best_q0}. 
Let us now consider the quantity $\imag q_0$ and try to fit it to a power-law 
Ansatz. We obtain $\imag q_0 \approx 26.4 m^{-4.45}$,
indicating that the true behavior might in fact be exponential. Thus, we
tried an exponential Ansatz $\imag q_0 = A \times B^m$. We 
obtain a very good fit including all the data: $A = 0.1177(39)$ and 
$B= 0.6188(26)$ with $\chi^2 = 0.019$ (1 degree of freedom (DF), confidence
level (level) = 89\%).\footnote{
  We have fitted the data shown in Table~\protect\ref{table_summary}; 
  the error bars are assigned to be one unit in the least significative digit. 
} 
The similarity between $B^{-1}= 1.6160(67)$ and the golden ratio 
$\tau= B_5^{1/2} = (1+\sqrt{5})/2 \approx 1.6180339887$ is striking. One 
may speculate in a possible connection between this appearance of $B_5$ and 
its role as the largest real isolated limiting point in the $m\to\infty$ 
limit. Thus, we state the conjecture  

\begin{conjecture}
\label{conj_even1}
The limiting curves for square-lattice strips with cylindrical boundary 
conditions and even width $m$ do not cross the real $q$-axis. Let $q_0(m)$ be 
the closest point of the limiting curve to the real $q$-axis. Then, 
$\real q_0(m)$ is a monotonically increasing function of $m$, and  
$\imag q_0(m) \sim B_5^{-m/2} = \tau^{-m}$ where $\tau = (1+\sqrt{5})/2$ is
the golden ratio. 
\end{conjecture}

Let us go back to the quantity $q_0$ for lattice strips with odd $m$. 
We can try to fit the data to the power-law Ansatz $A + B m^{-\Delta}$. 
In all the fits we impose a lower cutoff $m \geq m_{\rm min}$ to the
data and we study the effect of the cutoff on the estimates $A$, $B$ and 
$\Delta$ and on the $\chi^2$. As we increase $m_{\rm min}$ the estimate $A$ 
decreases from $3.12688(2)$ down to $3.02799(5)$; the estimate $-B$ increases
from $1.49601(4)$ to $2.8011(19)$; and the exponent $\Delta$ increases from
$0.76651(40)$ to $1.25578(42)$. The $\chi^2$ is very poor: $8.69 \times 10^6$ 
and $1.75\times 10^5$ for $m_{\rm min}=5$ and $7$ respectively. It seems  
that the estimate for $A$ tends to 3 from above. As we know that the true value
for $A$ should be bounded from above by $q_c=3$, we can try to fit the data 
to the 
Ansatz: $3-q_0 = B m^{-\Delta}$. Again, we do not find any stable fit: 
as $m_{\rm min}$ increases from $5$ to $11$, the estimate $B$ monotonicly
goes from $2.47978(4)$ to $5.3807(12)$, and the exponent $\Delta$ grows
from $1.28668(1)$ to $1.62265(9)$. Again the $\chi^2$ is poor: it 
goes from $1.72\times10^8$ for $m_{\rm min}=5$ to $5.22\times 10^5$ for
$m_{\rm min}=9$.  

We can play the same game with the quantity $\real q_0$ for the lattice strips
of even width $m$. The fits to the Ansatz $A + B m^{-\Delta}$ are similar
to the former case: as $m$ is increased, $A$ decreases from $3.22832(3)$ 
down to $3.02285(28)$; $-B$ increases from $2.95545(6)$ to $5.970(19)$; and 
$\Delta$ increases from $0.87657(4)$ to $1.4866(20)$. The $\chi^2$ is poor: 
$5.19 \times 10^6$ and $5.30\times 10^3$ for $m_{\rm min}=4$ and $6$ 
respectively. Again the estimate for $A$ seems to tend to 3 from above, 
so we can try 
the next Ansatz $3-\real q_0 = B m^{-\Delta}$. These fits are not very stable
either: as $m_{\rm min}$ increases from $6$ to $10$, $B$ monotonicly 
goes from $6.11369(19)$ to $8.989(12)$, and $\Delta$ grows
from $1.54057(2)$ to $1.71849(54)$. The $\chi^2$ is also poor: it 
goes from $7.02\times10^5$ for $m_{\rm min}=6$ to $8.45\times 10^3$ for
$m_{\rm min}=8$. 

Our data thus suggest that, 
within our numerical precision, $q_0({\rm sq}) = q_c({\rm sq}) = 3$. We can 
state this finding as a conjecture

\begin{conjecture}
\label{conj_q0}
For the zero-temperature square-lattice Potts antiferromagnet with 
cylindrical boundary conditions, $q_0=q_c=3$.
\end{conjecture}
 

The various limiting curves ${\cal B}$ are most similar for intermediate
values of $\real q$ (namely, $0.6\ltapprox \real q \ltapprox 2.4$).
The finite-size effects are more
apparent for $q$ close to $q=0$ and $q=q_c$. This is expected from what 
Baxter \cite{Baxter_87} found for the triangular lattice: the density of points
around $q=0$ and $q=q_c$ was smaller than for the other regions in the complex 
$q$-plane; thus, the convergence is expected to be slower around those two
points. 

In particular, we expect that eventually the rightmost branches of each
limiting curve (see Figure~\ref{Figure_sq_all}) will curve  
back and decrease towards the expected value $q_c=3$.
Indeed, this phenomenon of ``overshooting'' seems to be characteristic
of cylindric boundary conditions; comparison with the limiting
curves ${\cal B}$ found in Ref.~\cite{transfer1} for {\em free} boundary
conditions shows that in this case the approach towards the limiting
value $q_c=3$ is monotonic. A similar remark holds true for the Potts
antiferromagnet on the triangular lattice \cite{transfer3}.
On the other hand, the leftmost endpoints of the limiting curves ${\cal B}$ 
are expected to curve back and converge to  the point $q=0$ as in the 
triangular-lattice case.
 
We have also checked Conjectures 7.1, 7.2, and 7.3 of 
Ref.~\cite[Section 7]{transfer1}. In particular, they state that for a
square-lattice strip of width $L$ with free or cylindrical boundary conditions, 
\begin{enumerate}
\item At each Beraha number $q=B_2,\ldots,B_{L+1}$ there is at least one 
vanishing amplitude $\alpha_k(q)$;
\item For all $q=B_k$ with $k>L+1$ none 
of the amplitudes $\alpha_k(q)$ vanishes; and
\item $\det D(q) >0$ for all 
$q=B_k$ with $k>L+1$.
\end{enumerate}
We have shown that Conjectures 7.2 and 7.3 hold up to
$k=50$ for $L=9_{\rm P}$, and up to $k=22$ for 
$10_{\rm P} \leq L \leq 13_{\rm P}$.
In all cases, we have found empirically that the number of vanishing 
amplitudes decreases from $q=B_2$ to $q=B_{L+1}$. In this latter case there is 
only one vanishing amplitude and its corresponding eigenvalue is given by
$\lambda = (-1)^L$. We have checked that this is so for all $L\leq 13_{\rm P}$ 
and $L\leq 8_{\rm F}$.\footnote{
  We have only found one exception to this empirical behavior: for 
  $L=5_{\rm F}$ there are two eigenvalues $\lambda=-1$. 
}

\section*{Acknowledgments}

We wish to thank Dario Bini for supplying us the MPSolve 2.0 package
\cite{Bini_package} and for numerous discussions about its use, and Alan Sokal 
for many helpful conversations throughout the course of this work.
We are also grateful to a referee for suggesting the conjecture on
the asymptotic behavior of ${\rm Im} \, q_0$.
The authors' research was supported in part by
CICyT (Spain) grant AEN99-0990 (J.S.).

%
%

%
%

%
%
\begin{table}[t]
\centering
\scriptsize
\begin{tabular}{|l|l|l|l|}
\hline\hline
 Lattice & 3rd Zero & 4th Zero   & 5th Zero   \\
\hline\hline
$ 9_{\rm P}\times 9_{\rm F}$ &  2   &  2.620608391171  &  2.660519126718  \\
$ 9_{\rm P}\times 18_{\rm F}$ &  2   &  2.618033988527  &   \\
$ 9_{\rm P}\times 27_{\rm F}$ &  2   &  2.618033988750  &  2.791347005408  \\
$ 9_{\rm P}\times 36_{\rm F}$ &  2   &  2.618033988750  &   \\
$ 9_{\rm P}\times 45_{\rm F}$ &  2   &  2.618033988750  &  2.814819876147  \\
$ 9_{\rm P}\times 54_{\rm F}$ &  2   &  2.618033988750  &   \\
$ 9_{\rm P}\times 63_{\rm F}$ &  2   &  2.618033988750  &  2.824928798467  \\
$ 9_{\rm P}\times 72_{\rm F}$ &  2   &  2.618033988750  &   \\
$ 9_{\rm P}\times 81_{\rm F}$ &  2   &  2.618033988750  &  2.830572841537  \\
$ 9_{\rm P}\times 90_{\rm F}$ &  2   &  2.618033988750  &   \\
\hline
$ 10_{\rm P}\times 10_{\rm F}$ &  2.000000000000  &  2.614232442467  &   \\
$ 10_{\rm P}\times 20_{\rm F}$ &  2.000000000000  &  2.618033987905  &   \\
$ 10_{\rm P}\times 30_{\rm F}$ &  2.000000000000  &  2.618033988750  &   \\
$ 10_{\rm P}\times 40_{\rm F}$ &  2.000000000000  &  2.618033988750  &   \\
$ 10_{\rm P}\times 50_{\rm F}$ &  2.000000000000  &  2.618033988750  &   \\
$ 10_{\rm P}\times 60_{\rm F}$ &  2.000000000000  &  2.618033988750  &   \\
$ 10_{\rm P}\times 70_{\rm F}$ &  2.000000000000  &  2.618033988750  &   \\
$ 10_{\rm P}\times 80_{\rm F}$ &  2.000000000000  &  2.618033988750  &   \\
$ 10_{\rm P}\times 90_{\rm F}$ &  2.000000000000  &  2.618033988750  &   \\
$ 10_{\rm P}\times 100_{\rm F}$ &  2.000000000000  &  2.618033988750  &   \\
\hline
$ 11_{\rm P}\times 11_{\rm F}$ &  2   &  2.618035720465  &  2.731088786113  \\
$ 11_{\rm P}\times 22_{\rm F}$ &  2   &  2.618033988750  &   \\
$ 11_{\rm P}\times 33_{\rm F}$ &  2   &  2.618033988750  &  2.836691210598  \\
$ 11_{\rm P}\times 44_{\rm F}$ &  2   &  2.618033988750  &   \\
$ 11_{\rm P}\times 55_{\rm F}$ &  2   &  2.618033988750  &  2.857656644942  \\
$ 11_{\rm P}\times 66_{\rm F}$ &  2   &  2.618033988750  &   \\
$ 11_{\rm P}\times 77_{\rm F}$ &  2   &  2.618033988750  &  2.866767905070  \\
$ 11_{\rm P}\times 88_{\rm F}$ &  2   &  2.618033988750  &   \\
$ 11_{\rm P}\times 99_{\rm F}$ &  2   &  2.618033988750  &  2.871875414535  \\
$ 11_{\rm P}\times 110_{\rm F}$ &  2   &  2.618033988750  &   \\
\hline
$ 12_{\rm P}\times 12_{\rm F}$ &  2.000000000000  &  2.618032077436  &   \\
$ 12_{\rm P}\times 24_{\rm F}$ &  2.000000000000  &  2.618033988750  &   \\
$ 12_{\rm P}\times 36_{\rm F}$ &  2.000000000000  &  2.618033988750  &   \\
$ 12_{\rm P}\times 48_{\rm F}$ &  2.000000000000  &  2.618033988750  &   \\
$ 12_{\rm P}\times 60_{\rm F}$ &  2.000000000000  &  2.618033988750  &   \\
$ 12_{\rm P}\times 72_{\rm F}$ &  2.000000000000  &  2.618033988750  &   \\
$ 12_{\rm P}\times 84_{\rm F}$ &  2.000000000000  &  2.618033988750  &   \\
$ 12_{\rm P}\times 96_{\rm F}$ &  2.000000000000  &  2.618033988750  &   \\
\hline
$ 13_{\rm P}\times 13_{\rm F}$ &  2   &  2.618033988926  &  2.775231591227  \\
$ 13_{\rm P}\times 26_{\rm F}$ &  2   &  2.618033988750  &   \\
$ 13_{\rm P}\times 39_{\rm F}$ &  2   &  2.618033988750  &  2.867878041004  \\
$ 13_{\rm P}\times 52_{\rm F}$ &  2   &  2.618033988750  &   \\
$ 13_{\rm P}\times 65_{\rm F}$ &  2   &  2.618033988750  &  2.886773116380  \\
\hline
\hline
 Beraha &2   & 2.618033988750  &     \\
\hline
\end{tabular}
\caption{\protect\label{table_zeros_cyl}
   Real zeros of the chromatic polynomials of finite square-lattice strips
   with periodic boundary conditions in the transverse direction
   and free boundary conditions in the longitudinal direction,
   to 12 decimal places.
   A blank means that the zero in question is absent.
   The first two real zeros $q=0,1$ are exact on all lattices;
   the third real zero $q=2$ is exact on all lattices of odd width.
   ``Beraha'' indicates the Beraha numbers $B_4 = 2$ and
   $B_5 = (3+\sqrt{5})/2$.
}
\end{table}

\clearpage

%
%
\begin{table}
\small
\hspace*{-1.7cm}
\begin{tabular}{|c||c|c|c|c|c|c|c|c||c|c|}
\cline{2-11}
\multicolumn{1}{c||}{\mbox{}}&
\multicolumn{8}{|c||}{Eigenvalue-Crossing Curves ${\cal B}$} &
\multicolumn{2}{|c|}{Isolated Points}\\
\hline\hline
Lattice      & \# C & \# E & \# T & \# D & \# ER & $\min \real q$ & 
$q_0$                      & $\max \real q$      & \# RI& \# CI   \\
\hline\hline
$3_{\rm P}$  &   0  &    0 & 0    &   0  &   0   &                 &  
                           &                     &  3   & 0       \\
$4_{\rm P}$  &   3  &    8 & 0    &   1  &   0   & \phantom{$-$}0.709803 & 
   $[2.253370, 2.351688]$  &            2.995331 &  3   & 0       \\
$5_{\rm P}$  &   5  &   10 & 0    &   0  &   0   & \phantom{$-$}0.165021 & 
                  2.691684 &            2.691684 &  4   & 0       \\
$6_{\rm P}$  &   3  &   10 & 2    &   1  &   0   & $-0.131889$      & 
   $[2.608943, 2.613228]$  &            3.171192 &  3   & 1       \\
$7_{\rm P}$  &   7  &   16 & 2    &   0  &   0   & $-0.296250$      & 
                  2.788378 &            2.839016 &  4   & 1       \\
$8_{\rm P}$  &   6  &   16 & 4    &   0  &   0   & $-0.390864$      & 
$2.751531 \pm 0.002531\,i^*$&           3.211132 &  4   & 2        \\
$9_{\rm P}$  &   9  &   20 & 2    &   0  &   0   & $-0.446992$      &
$2.850572$                 &            3.011232 &  4   & 4        \\
$10_{\rm P}$ & $10^\dagger$&$22^\dagger$&$2^\dagger$&$0^\dagger$&
$0^\dagger$ & $-0.48079\phantom{0}$ &
$2.82813 \pm 0.00097\,i^*$.& 3.21691\phantom{0} & 4  & $6^\dagger$   \\
$11_{\rm P}$ & $9^\dagger$&$20^\dagger$&$2^\dagger$&$0^\dagger$& 
$0^\dagger$ & $-0.50104\phantom{0}$ &
$2.890093$                 & 3.09853\phantom{0} & 4  & $6^\dagger$    \\
$12_{\rm P}$ &      &      &      &      &       &      &
$2.87436 \pm 0.00037\,i^*$ &      & 4  & $6^\dagger$  \\
$13_{\rm P}$ &      &      &      &      &       &      &
$2.916189$          &      & 4  &  $6^\dagger$    \\
\hline\hline
\end{tabular}

\vspace{1cm}
\caption{\protect\label{table_summary}
   Summary of qualitative results for the eigenvalue-crossing curves $\scrb$
   and for the isolated limiting points of zeros.
   For each square-lattice strip considered in this paper,
   we give the number of connected components of $\scrb$ (\# C),
   the number of endpoints (\# E),
   the number of T points (\# T),
   the number of double points (\# D),
   and the number of enclosed regions (\# ER);
   we give the minimum value of $\real q$ on $\scrb$,
   the value(s) $q_0$ where $\scrb$ intersects the real axis
   (${}^*$ denotes an almost-crossing),
   and the maximum value of $\real q$ on $\scrb$.
   We also report the number of real isolated limiting points of zeros
   (\# RI) [which are always successive Beraha numbers $B_2$, $B_3$, \ldots]
   and the number of complex conjugate pairs of isolated limiting points
   (\# CI).
   The symbol $^\dagger$ indicates uncertain results. The results for 
   $L \leq 8_{\rm P}$ are included for comparison; they are taken from  
   Ref.~\protect\cite{transfer1}.
}
\end{table}

\clearpage

%
%
%
%

\clearpage
%
%
%
%
%

\clearpage
%
%
%
%
%

\clearpage
%
%
%
%
%

\clearpage

%
%

\clearpage
%
%

\clearpage

\end{document}